\documentclass[fleqn,usenatbib]{mnras}

\usepackage{newtxtext,newtxmath}
\usepackage[normalem]{ulem}
\hypersetup{draft}
\usepackage[T1]{fontenc}
\usepackage{ae,aecompl}

\usepackage{graphicx}	% Including figure files
\usepackage{amsmath}	% Advanced maths commands
\usepackage{amssymb}	% Extra maths symbols
\usepackage{subfig}
\newcommand{\higpus}{\textsc{HiGPUs}}
\newcommand{\msun}{\,\mathrm{M}_{\sun{}}}

\title[Pulsar jerks in globular clusters]{
Intermediate mass black holes in globular clusters: effects on jerks and jounces of millisecond pulsars
}

\author[F. Abbate et al.]{
Federico Abbate,$^{1,2}$\thanks{E-mail: f.abbate@campus.unimib.it}
Mario Spera,$^{1,3,4,5,6,7}$
Monica Colpi$^{1,8}$
\\
$^{1}$Dipartimento di Fisica `G. Occhialini', Universit\`a degli Studi Milano - Bicocca, Piazza della Scienza 3, Milano, Italy\\
$^{2}$INAF - Osservatorio Astronomico di Cagliari, Via della Scienza, I-09047 Selargius (CA), Italy\\
$^{3}$Dipartimento di Fisica e Astronomia `G. Galilei', University of Padova, Vicolo dell'Osservatorio 3, I--35122, Padova, Italy\\
$^{4}$INFN, Sezione di Padova, Via Marzolo 8, I--35131, Padova, Italy\\
$^{5}$Institut f\"ur Astro- und Teilchenphysik, Universit\"at Innsbruck, Technikerstrasse 25/8, A-6020, Innsbruck, Austria\\
$^{6}$ Department of Physics and Astronomy, Northwestern University, Evanston, IL 60208, USA\\
$^{7}$Center for Interdisciplinary Exploration and Research in Astrophysics (CIERA), Evanston, IL 60208, USA\\
$^{8}$ Istituto Nazionale di Fisica Nucleare, Sezione di Milano Bicocca, Piazza della Scienza 3, Milano, Italy\\
}

\date{Accepted XXX. Received YYY; in original form ZZZ}

\pubyear{2018}

\begin{document}
\label{firstpage}
\pagerange{\pageref{firstpage}--\pageref{lastpage}}
\maketitle

\begin{abstract}
Globular clusters may host intermediate mass black holes (IMBHs) at their centres. 
 Here we propose a new method for their identification using millisecond pulsars (MSPs) as probes.
We show  that measuring the first (jerk) and second (jounce) derivatives of the accelerations of an ensemble of MSPs will let us infer the presence of an IMBH in a globular cluster better than measuring
the sole accelerations.
We test this concept by simulating a set of star clusters with and without a central IMBH to extract the distributions of the stellar jerks and jounces. We then apply this technique to the ensemble of 
MSPs in the Galactic globular cluster 47 Tucanae.  Current timing observations  are insufficient to constrain the presence of an IMBH and can only be used to pose upper limits on its mass. But, with few more years of observations it will be possible to test for the presence of a central IMBH with mass smaller than $\sim$ 1000 M$_{\odot}$. 
We conclude that jerks and jounces help significantly in reducing the upper limit of the mass of IMBHs in Galactic globular clusters.

\end{abstract}


%%%%%%%%%%%%%%%%%%%%%%%%%%%%%%%%%%%%%%%%%%%%%%%%%%

%%%%%%%%%%%%%%%%% BODY OF PAPER %%%%%%%%%%%%%%%%%%

\section{Introduction}

%\subsection{Globular clusters, IMBHs and millisecond pulsars}

Globular clusters (GCs) play a crucial role in a plethora of astrophysical phenomena. For instance, they are unique environments to study the interplay between stellar dynamics and stellar evolution, they host multiple stellar populations, they are nurseries of high frequency gravitational-wave sources \citep{Benacquista13,Rodirguez18,Choksi18}, and some of them may host an intermediate mass black hole (IMBH) at their centre \citep{Miller2002}.
IMBHs are the black holes with mass between $10^2 -10^5$ M$_{\odot}$, which are considered the missing link between the less massive stellar black holes and the super-massive black holes found in the centre of galaxies  as they are often referred as seeds upon which  super-massive black holes have grown  \citep{Volonteri2010,Latif2016,Johnson2016,Mezcua2017}. Whether or not GCs host IMBHs has been matter of debate for a long time  \citep{Bahcall1975}. Dense stellar systems like GCs could be breeding grounds for the formation of IMBHs, via stellar collisions or gravitational encounters among stellar black holes followed by mergers \citep{Miller2002,PortegiesZwart2002,Gurkan2004,PortegiesZwart2004,Giersz2015}.
Furthermore, if bound to a black hole in a close binary, IMBHs are sources of gravitational waves, detectable with the next generation of interferometers as the Laser Interferometer Space Antenna and Einstein Telescope \citep{LISA2017,SathyaET2011,Amaro-Seoane2010,Gair2011}.
Limits on the coalescence rate of IMBHs with masses of $\sim 100\,\rm M_\odot$ have been recently posed by the Advanced LIGO and Advanced Virgo Collaboration \citep{Abbott2017}.

Two methods have been widely used to search for IMBHs in GCs. The first method relies upon the study of the dynamics of stars in clusters, which is affected by the presence of a central IMBH \citep{Bahcall1976, Gebhardt2005}. 
Through optical observations, it is possible to measure the projected position of stars on the sky, the line-of-sight velocity and their proper motion. Using this technique, past searches inferred only upper limits on the masses of hypothetical IMBHs in GCs, and made a few tentative detections \citep{Mclaughlin2006, Noyola2008, vanderMarel2010}.

The second method relies on X-ray and radio continuum emission from the IMBHs in an accretion state \citep{Maccarone2004, Maccarone2008, Maccarone2010}. The searches based on this technique were able to get only controversial upper limits on the mass of  central IMBHs \citep{ Pooley2006,Lu2011, Miller-Jones2012,Tremou2018}. Recently, evidence towards the existence of IMBHs in stellar clusters has been found through observations of a tidal disruption event in an extra-galactic stellar cluster \citep{Lin2018}.

An alternative method to search for IMBHs comes from millisecond pulsars (MSPs) which are commonly found in GCs \citep{Perera2017, Kiziltan2017b, Freire2017, Prager2017, Abbate2018}.
Thanks to their extremely stable rotational periods, measurements of the Doppler shift allow us to recover the acceleration felt by the MSP in the GC and its time derivatives. The use of these measurements to map the gravitational potential of the GC was anticipated by  \cite{Blandford1987} and the concept was further developed by \cite{Phinney1993}.

While the effects of a central IMBH on the acceleration of MSPs in a GC have been studied in the past \citep{Prager2017}, the effects on the first derivative of the acceleration, referred to as {\it jerk}, and the second derivative, generally referred to as {\it  jounce} or {\it snap}, have usually been neglected, despite prolonged timing of MSPs in GCs shows that jerks and jounces are measurable \citep{Freire2017, Perera2017, Liu2018}.

In this paper, we propose a new method of identifying IMBH candidates in Galactic GCs which involves the measurements of high-order time derivatives of the acceleration on an ensemble of MSPs. To this purpose, we carry out a suite of direct $N$-body simulations of star clusters where we compute self-consistently the high-order derivatives of accelerations, treating MSPs as test particles.
%after a direct analysis of jerks and jounces of stars in a suite of direct N-body simulations that can be used as laboratory for testing the dynamical effect that an embedded IMBH has on stars, and thus on MSPs. 
We show that the aforementioned derivatives contain crucial information on the mean field gravitational potential as well as on the coarse-grain effects caused by neighbouring stars, both of which are affected by the presence of an IMBH. 
We show that this method is sensible to IMBHs of lower mass than current standard methods.
We apply this technique on a synthetic star cluster to test its capability of detecting a central IMBH, as well as on the GC 47 Tucanae (also known as NGC 104, hereafter 47 Tuc). We also test how the results would improve with longer datasets of observations with both Parkes and MeerKAT radio telescopes.

In Section \ref{theory} we introduce the mathematical expressions of jerks and jounces, also affected by the central IMBH, and describe the $N$-Body simulations used to study our synthetic GC systems. In Section \ref{results} we present the results on the simulated clusters and on 47 Tuc. Section \ref{conclusion} 
summarises our results.

\section{Methods} \label{theory}
In this section we derive the mathematical expressions of jerks and jounces of stars in a self-gravitating stellar system described by a King model \citep{King1962}. We consider the contributions from the mean field, the nearest neighbouring stars and the IMBH. Furthermore, we account for modification in the stellar background
due to the presence of the black hole.

\subsection{Jerks}\label{section_jerks}

In a GC described by a King profile \citep{King1962}, a test star experiences the gravitational attraction from the mean field generated by all the stars in the cluster. Within a few core radii this acceleration 
is described by \citep{Freire2005}

\begin{equation}\label{acc_king}
{\bf a}(r) = -4 \pi G \rho_{\rm c} r_{\rm c}^3 \left[ \sinh^{-1} \left( \frac{r}{r_{\rm c}}\right) - \frac{r}{r_{\rm c} \sqrt{1 + (r/r_{\rm c})^2} } \right] \frac{{\bf{r}}}{r^{3}} =-|a(r)| \frac{\bf r}{r},
\end{equation}
where $\mathbf{r}$ is the distance from the centre of the cluster, $\rho_{\rm c}$ the central cluster density and  $r_{\rm c}$  the core radius. Equation (\ref{acc_king}) is computed integrating the density profile $\rho(r)$ over a spherical volume of radius $r$. Within a few core radii, we can approximate 
\begin{equation}
\rho(r) = \frac{\rho_c}{[1+(r/r_c)^2]^{3/2}}.
\end{equation}
%and multiplying it by the factor $-G/r^2$.

The jerk is computed by taking the time derivative of equation (\ref{acc_king})

\begin{equation}\label{jerk_king}
\dot{\bf{a}}_K (r)= -\frac{d|a(r)|}{dt} \frac{\bf r}{r} - |a(r)| \frac{\bf v}{r} + |a(r)| \frac{\bf{(v \cdot r) \, r}}{r^3},
\end{equation}
where $|a(r)|$ is the norm of the acceleration and the time derivative of the norm is

\begin{equation}
\label{jerk_king_mod}
\frac{d|a(r)|}{dt}= \, -2 \frac{v |a(r)|}{r}  +4\pi G v \rho_{\rm c}  \left(\frac{1}{1 + (r/r_{\rm c})^2}\right)^{\frac{3}{2}}
\end{equation} 
with $v$ the norm of the velocity. 

In the case of GCs, which are collisional stellar systems, jerks are also heavily influenced by neighbouring stars and  can be as large as those from the mean field (\citealt{Prager2017}). Specifically, \cite{Prager2017}  found that the jerk caused by the coarse-grain nature of  stellar interactions is distributed with the following probability distribution:

\begin{equation}
P(\dot{\bf a})=\frac{1}{\pi^2} \frac{\dot a_0}{(\dot a^2 + \dot a_0^2)^2},
\end{equation}
where $\dot a_0$ is the characteristic jerk given by 
\begin{equation}
\dot a_0= \frac{2\pi \xi}{3} G \langle m \rangle \sigma n,
\end{equation}
where $\xi \simeq 3.04$ is a numerical constant, $\langle m \rangle$ is the average mass of the stars, $\sigma$ is
the velocity dispersion, and $n$ is the number density of the stars. The distribution of jerks projected along the line of sight, $\dot{a_l}$ is a Lorentzian distribution

\begin{equation}
P(\dot{a_l})=\frac{1}{\pi} \frac{\dot a_0}{\dot a_l^2 + \dot a_0^2}.
\end{equation}

If a central IMBH is present in the cluster, the jerk of a test star is affected by the central point mass $M$ 

\begin{equation}\label{jerk_general}
\dot{\bf{a}}_M = - G M \left( \frac{\bf{v}}{r^3} - 3 \frac{\bf{(v \cdot r) \, r}}{r^5} \right),
\end{equation}
where $\bf r$ the distance to the source $M$ and ${\bf v}$ is the relative velocity. Additionally, the IMBH creates a
 stellar over-density with radial profile with slope $-1.55$ \citep{Baumgardt2004b}. 
 The jerk produced by this over-density 
 %is found by calculating the time derivative of the acceleration by solving Newton's equation. The final expression has the form
 takes the form
\begin{equation} \label{jerks_cusp}
\dot{\bf a}_{\rm cusp} =
\begin{cases}
-\frac{4\pi G}{1.45} r_{\rm i}^{1.55} \rho_{\rm i} \left(  \frac{\bf v}{r^{1.55}} - 1.55 \frac{(\bf v \cdot r) r}{r^{3.55}} \right) \text{for } r<r_{\rm i}\\
-\frac{4\pi G}{1.45} r_{\rm i}^{3} \rho_{\rm i} \left(  \frac{\bf v}{r^{3}} - 3 \frac{(\bf v \cdot r) r}{r^{5}} \right) \text{for } r>r_{\rm i}\\
\end{cases}
\end{equation}
where $r_{\rm i}$ is the IMBH influence radius defined as $r_{\rm i} = GM /\sigma_{\rm c}^2$, $\rho_{\rm i}$ is the density at this radius, and $\sigma_{\rm c}$ the one-dimensional core stellar velocity dispersion \citep{Baumgardt2004b}. 

The cusp and the increase of the stellar velocity caused by the
presence of the IMBH affects also the rate of close encounters between stars. For this reason, even the jerk caused by the nearest neighbours is influenced, but a statistical description of this effect is not available yet. However, this effect is taken into account in our $N$-body simulations, because the integration algorithm self-consistently calculates all jerks, at each integration step.

A comparison between the mean field jerks derived using equations (\ref{jerk_king}), (\ref{jerk_general}) and (\ref{jerks_cusp}) and the ones estimated numerically from our  simulations is shown in Figure \ref{jerk_analytic} and commented in Appendix \ref{appendix}. 

\subsection{Jounces}\label{section_jounces}

In the case of GC described by a King profile, the jounce due to the mean field gravitational potential is given by

\begin{equation} \label{jounce_king}
\begin{split}
\ddot{\bf a}_K = -\frac{d^2 |a(r)|}{dt^2} \frac{\bf r}{r} - 2 \frac{d |a(r)|}{dt} \frac{\bf v}{r} + 2 \frac{d |a(r)|}{dt} \frac{\bf( v \cdot r) r}{r^3} +
\\
+ 5 |a(r)| \frac{\bf( v \cdot r) v}{r^3} - 3|a(r)| \frac{{\bf (v \cdot r )}^2 {\bf r}}{r^5},
\end{split}
\end{equation}
where $|a(r)|$ is defined in equation (\ref{acc_king}), $d |a(r)|/dt$ is given by equation (\ref{jerk_king_mod}) and $d^2 |a(r)|/dt^2$ is

\begin{equation}
\begin{split}
\frac{d^2 |a(r)|}{dt^2} = -\frac{d |a(r)|}{dt} \frac{|a(r)|}{v}\, - \,4 \frac{d |a(r)|}{dt} \frac{v}{r}\, - \,2\frac{|a(r)| v^2}{r^2}\, -\\
- \, \frac{4\pi G v^2}{r} \rho_{\rm c}\left[ \frac{1}{(1+(r/r_{\rm c})^2)^{3/2}} - \frac{3}{(1+(r/r_{\rm c})^2)^{5/2}} \right].
\end{split}
\end{equation}
Also for the jounce, the contribution from neighbouring stars plays a very important role. \cite{Phinney1993} shows that this contribution can be much larger than the one from the mean field.

Similarly to the jerks, the presence of an IMBH  influences jounces in two ways.  It contributes directly as a central point mass

\begin{equation}
\ddot{\bf a}_M= GM\left( - 2 a \frac{\bf r}{r^4}  - 6 \frac{(\bf v \cdot r) v}{r^5} -3\frac{v^2 {\bf r}}{r^5} +15 \frac{{\bf (v \cdot r )}^2 {\bf r}}{r^7}  \right),  
\label{jounce_general}
\end{equation}
and through the cusp over-density. At radial distances $r< r_{\rm i}$ the contribution reads

\begin{equation} \label{jounce_cusp_1}
\begin{aligned}
\ddot{\bf a}_{\rm cusp}= -\frac{4\pi G}{1.45} r_{\rm i}^{1.55} \rho_{\rm i} \left(-0.45 \frac{a {\bf r}}{r^{2.55}} -3.1 \frac{\bf (v \cdot r) v}{r^{3.55}} - \right. \\ \left. - 1.55 \frac{v^2 {\bf r}}{r^{3.55}} + 5.5 \frac{\bf (v\cdot r)^2 r}{r^{5.55}} \right),
\end{aligned}
\end{equation}
and for $r>r_{\rm i}$ as
\begin{equation}\label{jounce_cusp_2}
\begin{aligned}
\ddot{\bf a}_{\rm cusp}= -\frac{4\pi G}{1.45} r_{\rm i}^{3} \rho_{\rm i} \left(-2 \frac{a {\bf r}}{r^{4}} -6 \frac{\bf (v \cdot r) v}{r^{5}} - 5 \frac{v^2 {\bf r}}{r^{5}} + \right. \\ \left. + 15 \frac{\bf (v\cdot r)^2 r}{r^{7}} \right).
\end{aligned}
\end{equation}
Also the jounces caused by nearest neighbours are influenced by the presence of an IMBH but a statistical description of this effect is still missing. It is worth noting that we calculate self-consistently the values of jounces in our $N$-body simulations.

A comparison between the mean field jounces derived with the above equations (\ref{jounce_king}), (\ref{jounce_general}), (\ref{jounce_cusp_1}) and (\ref{jounce_cusp_2}) and the ones estimated numerically from simulations is shown in Figure \ref{jounce_analytic}, and commented in Appendix \ref{appendix}.

\subsection{Pulsar timing}\label{pulsar_timing}

Observations of radio pulsars provide us with precise time of arrivals (ToAs) of the pulses. Measuring the ToAs over different observations allows us to estimate the pulsars' rotational period $P$ and the time derivatives $\dot P$, $\ddot P$ and $\dddot P$, which require different timescales to be measured with high accuracy. While a single observation is enough to estimate the rotational period, around one year of observing time might be necessary to measure accurately the first derivative of the period for an MSP. Regular observations over many years are necessary to measure the second and third derivative. The first derivative of the period is mostly influenced by the line-of-sight acceleration of the MSP, whereas the second and third derivative are mainly influenced by jerks and jounces, respectively.

The relation between $\dot P$ and the acceleration can be written in the following way \citep{Phinney1993}

\begin{equation}
\left( \frac{\dot P}{P}\right)_{\rm meas} = \left( \frac{\dot P}{P}\right)_{\rm int} + \frac{a_{\rm c}}{c} +\frac{a_{\rm g}}{c} +\frac{\mu^2 D}{c}, \label{Pdot_p_equation}
\end{equation}
where $\left({\dot P}/P\right)_{\rm int}$ is the intrinsic spin-down of the pulsar caused by magnetic breaking of the neutron star \citep{Lorimer2005}, $ a_{\rm c}$ is the acceleration due to the gravitational potential of the cluster along the line of sight, $a_{\rm g}$ is the acceleration due to the Galactic potential along the line of sight, $\mu^2 D$ accounts the Shklovskii effect \citep{Shklovskii1970}, $\mu$ is the proper motion of the pulsars, $D$ is the distance of the cluster to the Sun, and $c$ is the speed of light.

\cite{Prager2017} and \cite{Abbate2018} showed that both the acceleration due to the Galactic potential and the Shklovskii effect have typical values of $10^{-11}-10^{-10}$ m s$^{-2}$. This means that they are negligible if compared with the acceleration due to the cluster potential, $a_{\rm c}$, which is usually $\sim 10^{-9}-10^{-8}$ m s$^{-2}$. The intrinsic spin-down depends on the values of the surface magnetic field. Its average value can be estimated by looking at MSPs in the Galactic disk, where the local acceleration is usually very small. Taking into account the population of Galactic MSPs \citep{Manchester2005}, \cite{Abbate2018} find that the contribution on the acceleration due to the intrinsic spin down is of the order of $10^{-9}$ m s$^{-2}$, which is of the same order of magnitude as $a_{\rm c}$. Therefore, it is very hard to disentangle the effects of the cluster acceleration from the intrinsic spin down. Any work focused on  measuring the acceleration in a GC from $\dot P$ will have large uncertainties due to the unknown intrinsic spin-down. 

%binary pulsars
We can measure directly the acceleration of an MSP only if it is a member of a binary system. For these systems it is possible to track the orbital period $P_{\rm b}$ and its derivative $\dot P_{\rm b}$ that is affected by the acceleration of the cluster due to the Doppler effect. The equation that describes the evolution of the orbital period is similar to equation (\ref{Pdot_p_equation}). Here the intrinsic orbital period derivative can be caused by the shrinking of the orbit due to the emission of gravitational waves or by other effects like mass loss from the companion. The change of the orbital period by gravitational wave emission can be estimated with the prescriptions of \cite{Damour1991} and is usually small compared with the acceleration of the GC. For the pulsars in 47 Tuc, the gravitational wave induced accelerations are of the order of $10^{-11}-10^{-10}$ m s$^{-2}$ \citep{Freire2017}. For binary pulsars with a white dwarf companion 
mass losses give negligible contributions, while for binaries with a low-mass main sequence star, corrections due to mass loss can be important and must be analysed case by case \citep{Freire2017}. However, for binary pulsars, the measure of the orbital period derivative is in general dominated by the acceleration of the GC.

The scenario is different when we consider the second derivative of the period which is related to jerks. The equation for the jerks is

\begin{equation}
\left(\frac{\ddot P}{P}\right)_{\rm meas} = \frac{ \dot{\rm a}_{\rm c}}{c} + \left(\frac{\ddot P}{P}\right)_{\rm int},\label{Pdotdot_p_equation}
\end{equation}
where ${\dot{\rm a}_{\rm c}}$ is the jerk due to the GC potential projected along the line of sight. We discuss later the effects resulting from the terms in equation (\ref{Pdot_p_equation}) which are neglected in equation (\ref{Pdotdot_p_equation}).

The contribution from the intrinsic spin-down due to magnetic dipole breaking
can be estimated following $\dot \omega_{\rm int}= -K \omega_{\rm int}^n$ \citep{Lorimer2005}, 
where $\omega=2\pi/P$, $K$ is a constant and $n$ is the breaking index assumed to be equal to 3.
This leads to a second time derivative for the period

\begin{equation}\label{intr_ddotp}
\left(\frac{\ddot P}{P}\right)_{\rm int} = (2-n) \left(\frac{\dot P}{P}\right)_{\rm int}^2.
\end{equation}

From the estimate of $(\dot P/P)_{\rm int}$ we find that the contribution of the intrinsic spin down on the jerk is of the order of $10^{-27}$ m s$^{-3}$. This is completely negligible when compared to the jerk due to the GC which is $\sim 10^{-18} - 10^{-19}$ m s$^{-3}$. This means that a measure of the second derivative of the period of a pulsar in a cluster corresponds to a direct measure of the jerk of the star.

Furthermore, we have a similar scenario for jounces. The equation for the third period derivative reads 

\begin{equation}
\left(\frac{\dddot P}{P}\right)_{\rm meas} = \frac{ \ddot{\rm a}_{\rm c}}{c} + \left(\frac{\dddot P}{P}\right)_{\rm int},
\end{equation}
where ${\ddot{\rm a}_{\rm c}}$ is the line-of-sight jounce due to the GC potential.
Estimating the contribution of the intrinsic spin-down from the time derivative of equation (\ref{intr_ddotp}), we find a value of $\sim 10^{-44}$ m s$^{-4}$. The latter contribution is negligible compared to the jounce due to the GC, which is $\sim 10^{-28}$ m s$^{-4}$. 

%other sources of errors (solar system ephemeris)

There are also other untreated sources of errors that affect the second and third period derivatives of MSPs. For instance, to compensate for all the delays caused by the Earth's motion, the ToAs of pulses are always referred to the Solar system barycentre. An imprecise determination of the barycentre may lead to inaccurate period derivatives \citep {Champion2010}. Other sources of errors like the effects of the  velocity both on the plane of the sky and on the line of sight have been discussed in a paper by \cite{Liu2018}.

Quantifying the effect of different sources of errors in the estimation of the second and third time derivative of the period is quite uncertain. As a reference for these uncertainties, we defer to  \cite{Perera2017} and \cite{Freire2017}. Converting the measured period derivatives to acceleration derivatives using the formulae written above, we find that we can measure jerks up to a precision of $\sim 10^{-21}$ m s$^{-3}$ and jounces up to $\sim 10^{-29}$ m s$^{-4}$. In Section \ref{BH_pulsars} we will investigate whether this level of accuracy is enough to quantify the effects of a central IMBH.

%Add discussion on binaries (center of mass)
Furthermore, a large number of pulsars in GCs are members of binary systems. The presence of a companion strongly affects the acceleration, jerk and jounce of the pulsar. However, if a binary system can be observed over a large number of orbits, it is possible to measure all the Keplerian parameters of the orbit and move the reference starting point of the pulses to the centre of mass of the system. In this way, we can correct the measures of accelerations, jerks and jounces for the effect of binarity.
If the binary period is longer than the tens of years of the MSP observations, we cannot correct for these effects. However, in an environment as dense as a GC, bound systems with orbital periods of tens of years are highly unstable and tend to be disrupted by dynamical interactions. Therefore, we assume that no MSP can be part of a binary with an orbital period that is so long.

\subsection{Simulations}\label{simulations}

We run a set of high-precision $N$-body simulations of star clusters by means of a new version of the direct summation $N$-body code \higpus{} \citep{Capuzzo-Dolcetta2013,dolcetta2013b}.  The \higpus{} code implements a Hermite 6th order integration method \citep{nitadori2008}, which uses accelerations, jerks and jounces to advance the positions and velocities of stars over time. The method is implemented using block time steps \citep{aarseth2003} and written with MPI and OpenCL, to  run on different and parallel computing architectures. We also implemented the new AVX-512 instructions to effectively run on the last generation Intel Central Processing Units as well as on Xeon Phi (Knights Landing) coprocessors.

We use the \higpus{} code to evolve a set of star clusters. Each star cluster is composed of $N=90,000$ stars, whose masses are distributed according to a \citet{kroupa2001} mass function with minimum mass $0.1\msun{}$ and maximum mass $2\msun{}$. \footnote{We use $2\msun{}$ as maximum mass of the mass function to mimic the stellar population of a quite old cluster.} 
The initial positions and velocities of stars are sampled from a \cite{king1966} density profile with central dimensionless potential W$_0=7$ and a core radius $r_{\rm c} = 0.16\,$pc, similar to the present-day distribution of stars in the GC Terzan 5. We evolved 6 different $N$-body models of the same star cluster, each one containing a central IMBH with mass $(0, 100, 500, 1000, 5000, 10000)$ $\msun{}$, respectively. The IMBH is initially placed at the centre of the star cluster with zero velocity, but it can wander during the $N$-body simulation, since it is treated as a real $N$-body particle. Furthermore, to minimise statistical fluctuations in our results, we evolve 10 different random realisations of the same star cluster. In our simulations, we do not use a softening parameter to artificially smooth out strong gravitational encounters, we do not include primordial binaries and we do not take into account stellar evolution calculations. 
All the star clusters start in virial equilibrium with the IMBH and have been evolved for $\sim 50$ Myr, to ensure that virial equilibrium is preserved and all systems are dynamically stable. The relative variation of the total energy (angular momentum) of each star cluster is kept below  $10^{-7}$ ($10^{-8}$) for all the simulations.

We also run a set of simulations reproducing the properties of 47 Tuc. The parameters are taken from \cite{Bellini2017}: the core radius is set to $r_{\rm c}= 0.58$ pc and the concentration parameter to $C=1.91$. Each $N$-body particle represents a single star and the number of particles in the simulations is chosen in order to obtain a total mass of $8.4 \times 10^5 \msun{}$. We evolved 80 different $N$-body models of the same star cluster each with an IMBH of different mass chosen randomly between 500 and 8000 $\msun{}$. We focus on this GC as it contains 25 known MSPs \citep{Freire2017} and has been subjected to various IMBH searches in the past with negative results \citep{Mclaughlin2006,Lu2011,Abbate2018}. There has been a claim of an IMBH of 2200 M$_{\odot}$ using the acceleration data from pulsars \citep{Kiziltan2017a}. However, this claim has not received further support from later studies on the same pulsar dataset \citep{Freire2017,Abbate2018}. 
As around 40 percent of these MSPs are isolated, we do not have accurate estimates of the acceleration for them (see section \ref{pulsar_timing}). This makes the jerks all the more useful when searching for an IMBH in 47 Tuc.

%Another cluster where this technique could be applied is Terzan 5 which contains 38 known pulsars\footnote{Full list of known pulsars in globular clusters available at: http://www.naic.edu/~pfreire/GCpsr.html}.
%However, we exclude this cluster from the analysis as it is highly obscured and difficult to observe optically. Because of this we have no clear indication on the density profile and core radius from this cluster independently from MSP data \citep{Prager2017}.

\section{Results}\label{results}

\begin{figure} 
\centering
\includegraphics[width=\columnwidth]{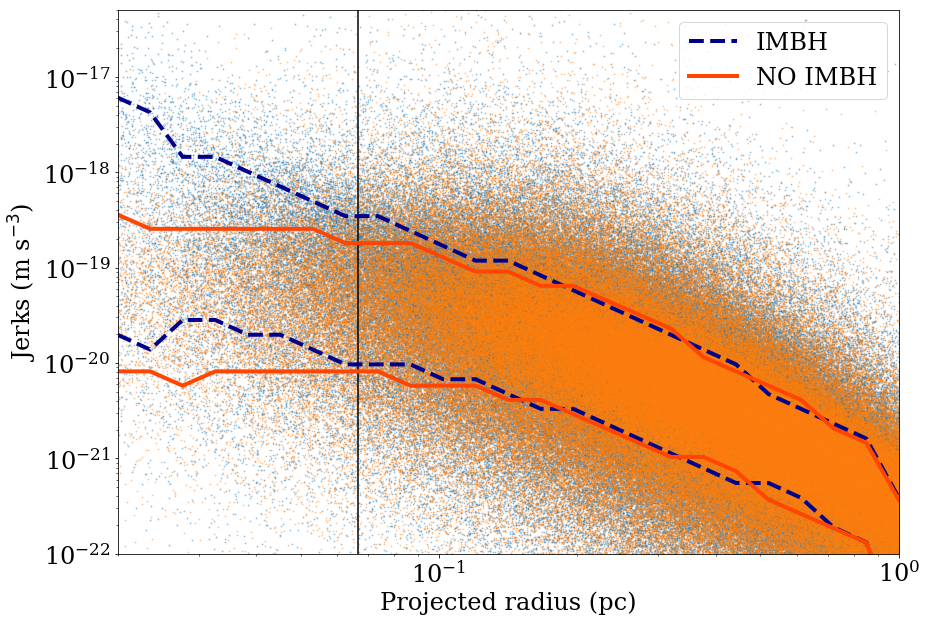}
\caption{Line-of-sight jerks of the stars from the simulations as a function of the projected distance from the cluster centre in presence of a black hole (blue dots) and without (orange dots). The IMBH mass is $1000 \, \rm{M}_{\odot}$.  The dashed blue and the solid red lines are the 1$\sigma$ interval of the distribution of the jerks measured over 25 radial bins. The vertical line is the influence radius of the IMBH. 
}
\label{jerk_total}
\end{figure}

Figure \ref{jerk_total} shows the line-of-sight jerks, and their 1$\sigma$ interval, as a function of the distance projected along the plane of the sky from the cluster centre. We compare the results from our $N$-body simulations of a star cluster with an IMBH of $1000\msun{}$ and the same star cluster without a central IMBH. 
The line of sight is chosen to correspond to the $z$ axis of the simulations. The 1$\sigma$ interval is computed by dividing the projected radius in 25 bins and measuring the 16 and 84 percentile of the jerks of the stars in each bin. The radius up to which the central black hole dominates the dynamics of the system is the influence radius. It is the radius at which the velocity of the Keplerian orbit around the black hole is the same as the stellar velocity dispersion in the core \citep{Baumgardt2004b}.
The effect of the IMBH is clearly visible inside its influence radius and extends further out. All the stars that passed inside the influence radius are affected either directly or indirectly by the IMBH and keep the memory of this interaction even when they leave the influence radius. This way the dynamical effects can extend farther outside.

%mean field and nearest neighbours
\begin{figure*}
\centering
%\begin{subfigure}{.45\textwidth}
\subfloat {\includegraphics[width=\columnwidth, height=0.3\textheight]{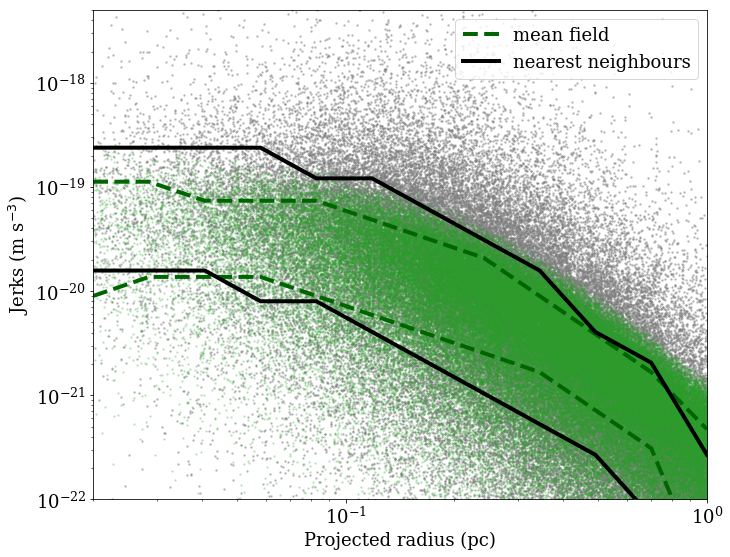}} %\begin{subfigure}{.45\textwidth}
\subfloat{\includegraphics[width=\columnwidth, height=0.3\textheight]{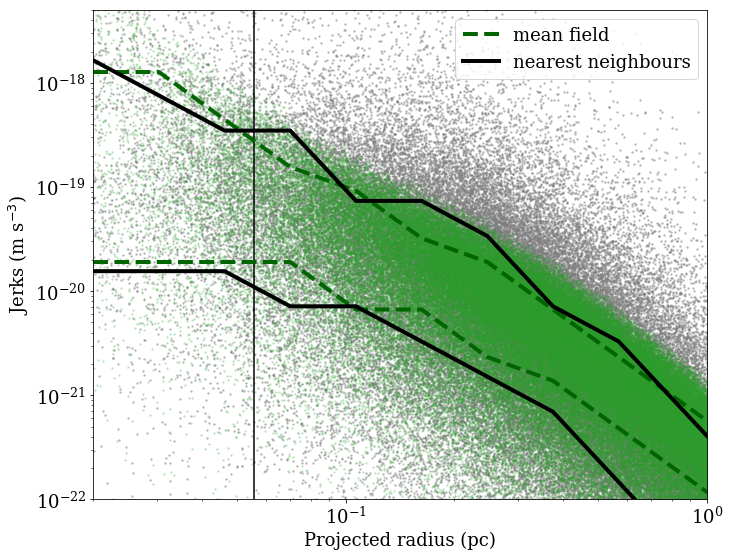} }%\quad

\caption{Line-of-sight jerks of the stars from the simulations as a function of the projected distance from the cluster centre in the case without the IMBH (left panel) and with an IMBH of $1000 \,\rm {M}_{\odot}$ (right panel).  Jerks due to the mean field are plotted with green dots and jerks due to the nearest neighbours are plotted with  grey dots. The dashed green and the solid black lines are the 1$\sigma$ interval of the distribution of the mean field jerks and of the nearest neighbour jerks measured over 12 radial bins. The vertical line shows the influence radius of the IMBH.}
\label{jerk_mf_nn}
\end{figure*}

The IMBH influences both the mean field and the nearest neighbour contribution to the jerks as seen in Figure \ref{jerk_mf_nn}. The mean field jerk is measured using equation (\ref{jerk_king}), while the nearest neighbours jerk is taken as the difference between the measured jerk and the mean field one. We find that the presence of the IMBH does not change the ratio of the two jerks, but affects their norms. 
The enhanced gravitational potential and stellar density within the influence radius lead to the increase of the mean field jerks. 
Likewise, the enhancement  in the local velocity dispersion and stellar density lead to the increase of the nearest neighbour jerks, as shown in Figure \ref{jerk_mf_nn}.

\begin{figure} 
\centering
\includegraphics[width=\columnwidth]{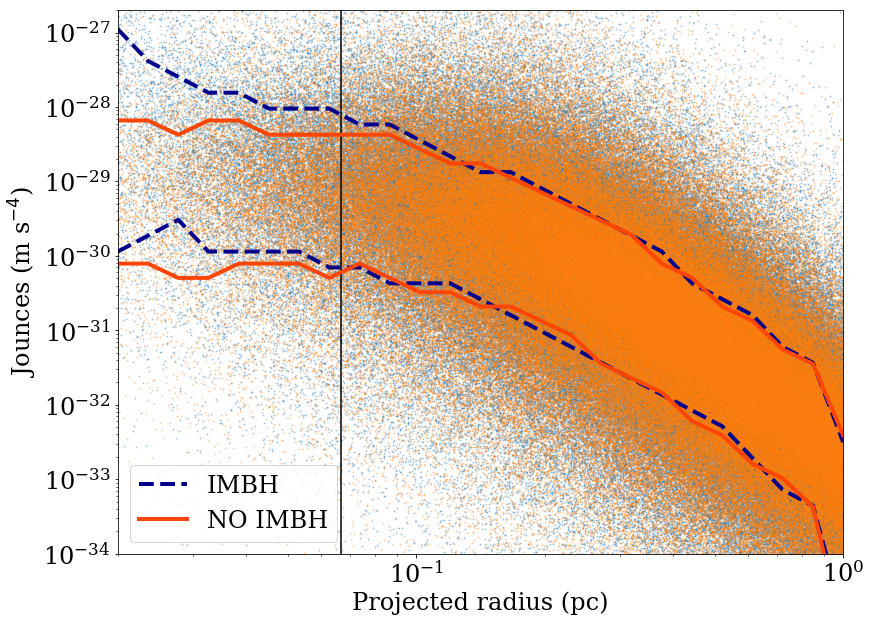}
\caption{Line-of-sight jounces of the stars from the simulations as a function of the projected distance from the cluster centre in  presence of an IMBH (blue dots) and without (orange dots). The black hole mass is $ 1000 \rm{M}_{\odot}$. The dashed blue and the solid red lines are the 1$\sigma$ interval of the distribution of the jounce measured over 25 radial bins. The vertical line is the influence radius of the black hole.}
\label{jounce_total}
\end{figure}

\begin{figure*}
\centering
%\begin{subfigure}{.48\textwidth}
\subfloat{\includegraphics[width=\columnwidth,height=0.3\textheight]{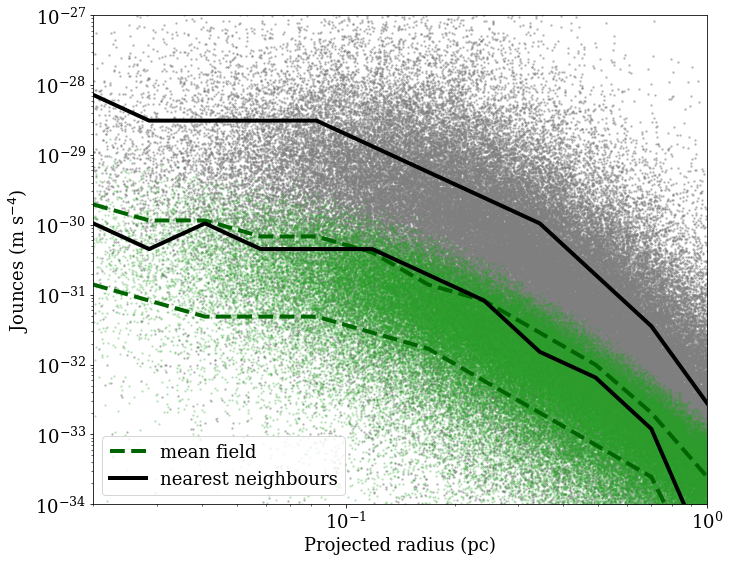} }%\quad
%\end{subfigure}
%\begin{subfigure}{.48\textwidth}
\subfloat{\includegraphics[width=\columnwidth,height=0.3\textheight]{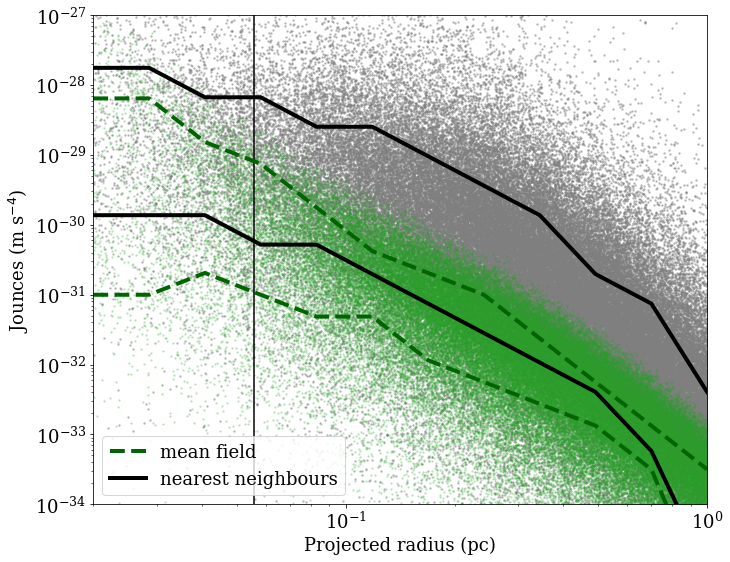} }%\quad
%\end{subfigure}
\caption{Line-of-sight jounces of the stars from the simulations as a function of the projected distance from the cluster centre  in the case without a black hole (left panel) and with a black hole of $1000 \, \rm{M}_{\odot}$ (right panel). The jounces due to the mean field are plotted with the green dots and the jounces due to the nearest neighbours are plotted with the grey dots. The dashed green and the solid black lines are the 1$\sigma$ interval of the distribution of the mean field jounces and of the nearest neighbour jounces measured over 12 radial bins. The vertical line shows the influence radius of the IMBH.
}\label{jounce_mf_nn}
\end{figure*}

Jounces projected along the line of sight and their 1$\sigma$ interval from the simulations are plotted in Figure \ref{jounce_total} for the case with an IMBH of 1000 M$_{\odot}$ and without. To create this plot we use the same prescriptions as in the case of jerks (Figure \ref{jerk_total}). The IMBH leads to an increase of the jounces, but less pronounced compared to that for the jerks.
If we disentangle the two contributions,  jounces from the mean field are affected the most but the contribution of the nearest neighbours is still dominant,
%jounces from the nearest neighbours are affected mostly, 
as shown in Figure \ref{jounce_mf_nn}.

\subsection{Detecting a black hole with pulsars}\label{BH_pulsars}

%In this section, we use MSPs as test particles for measuring jerks and jounces through their rotational period derivatives. \red{Beh, noi pero non usiamo i rotational period derivatives per fare questo studio dalle N-bosy simulations, giusto? Puoi aggiungere qui un po' di descrizione della metodologia? Cioe', noi abbiamo in realta' mappato le differenze nel piano jerks-radius nei casi di clusters con BHs e clusters senza BHs. E l'abbiamo fatto in funzione delle diverse masse di BHs. Quello che volgiamo vedere e' se siamo sensibili, e fino a che punto, sulle masse dei possibili IMBH. Puoi aggiungere qualche frasetta a riguardo?}

\begin{figure*}
\centering
%\begin{subfigure}{.45\textwidth}
\subfloat{\includegraphics[height=0.29\textheight]{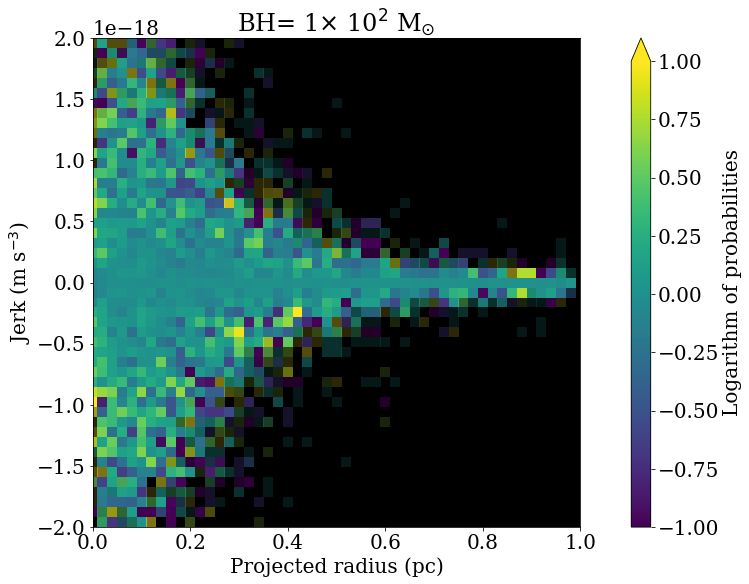} }%\quad
%\end{subfigure}
%\begin{subfigure}{.45\textwidth}
\subfloat{\includegraphics[height=0.29\textheight]{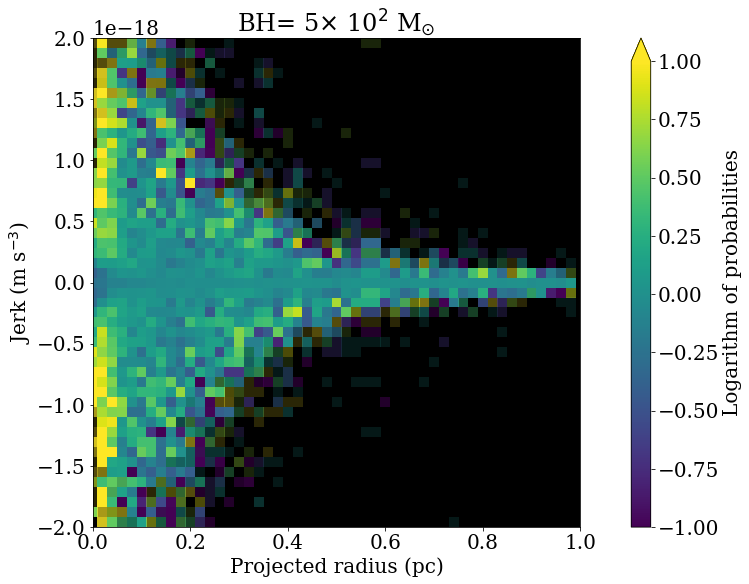} }%\quad
%\end{subfigure}

%\begin{subfigure}{.45\textwidth}
\subfloat{\includegraphics[height=0.29\textheight]{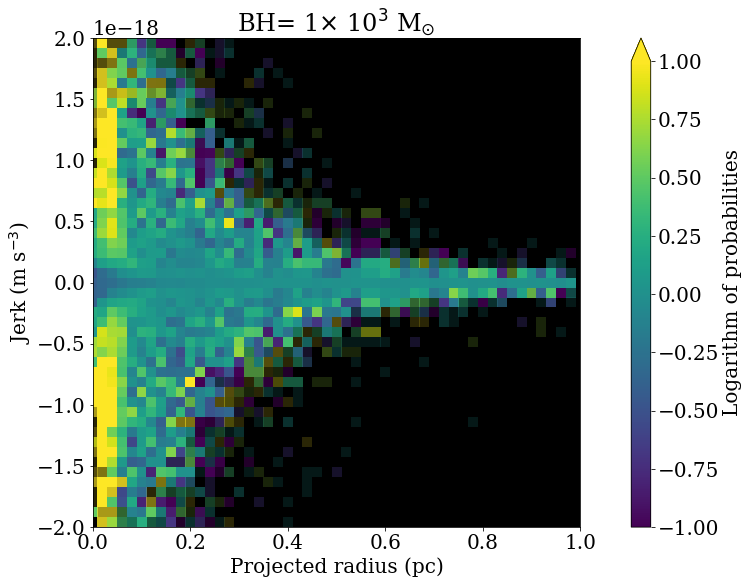}} %\quad
%\end{subfigure}
%\begin{subfigure}{.45\textwidth}
\subfloat{\includegraphics[height=0.29\textheight]{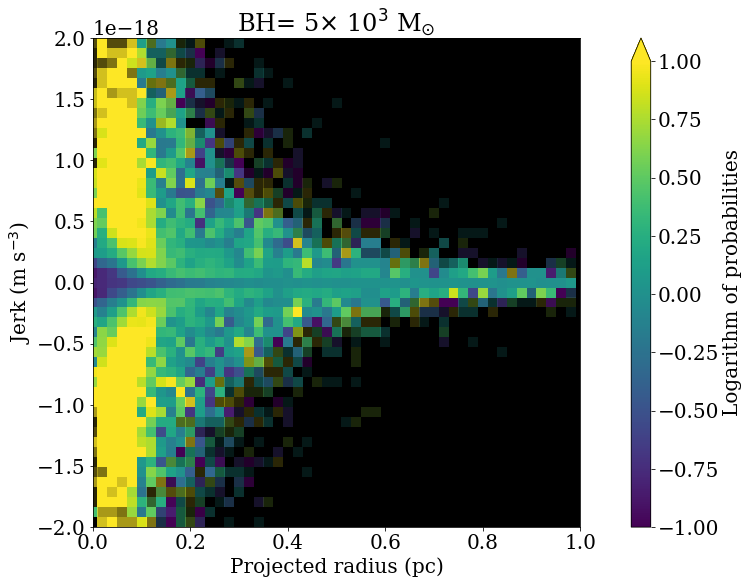}} %\quad
%\end{subfigure}

%\begin{subfigure}{.45\textwidth}
\subfloat{\includegraphics[height=0.29\textheight]{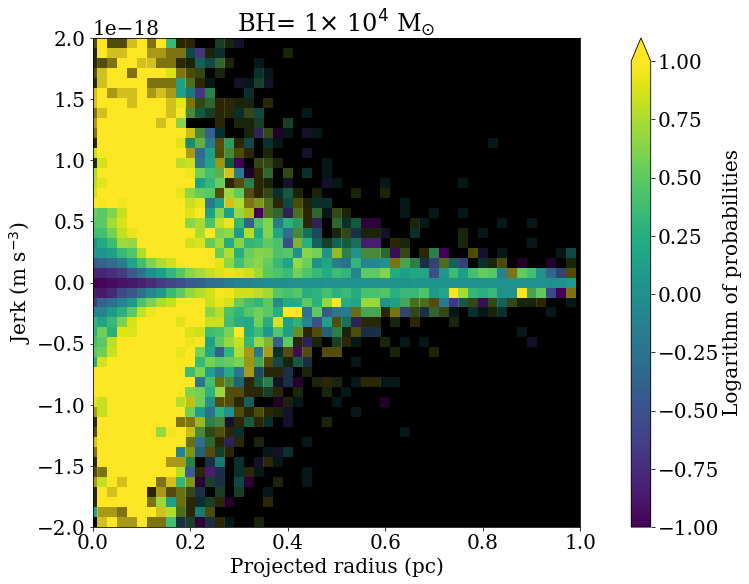}} %\quad
%\end{subfigure}
%\begin{subfigure}{.45\textwidth}
%\subfloat{\includegraphics[height=0.29\textheight]{new_jerk_prob_alpha_5e4.png}} %\quad
%\end{subfigure}

\caption{Probability of finding an MSP in a given projected radius bin with a given jerk in presence of an IMBH over the probability in the absence of it. The transparency of each pixel qualitatively shows the statistics of the value. Darker pixels have larger uncertainties. Pulsars found in a yellow square indicate that a black hole is likely present, while pulsars in blue square indicate that it is more likely that there is no black hole. }\label{jerk_prob_maps}
\end{figure*}

\begin{figure*}
\centering
%\begin{subfigure}{.45\textwidth}
\subfloat{\includegraphics[height=0.29\textheight]{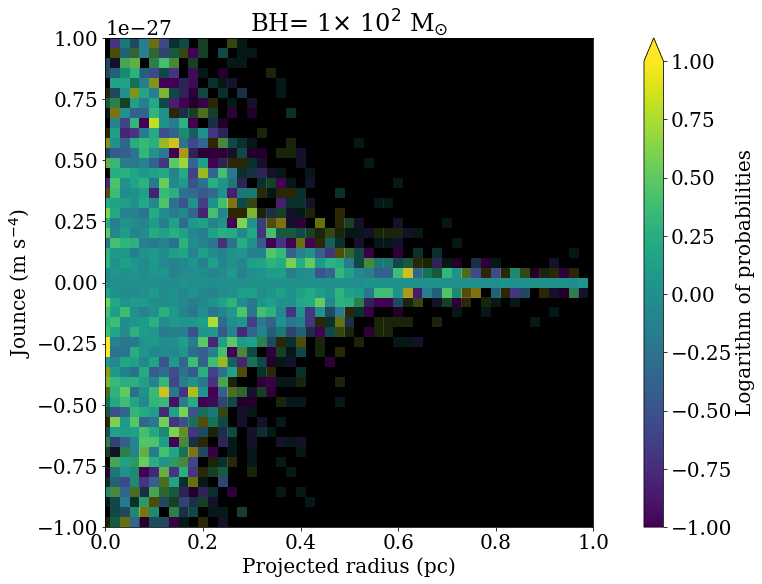}} %\quad
%\end{subfigure}
%\begin{subfigure}{.45\textwidth}
\subfloat{\includegraphics[height=0.29\textheight]{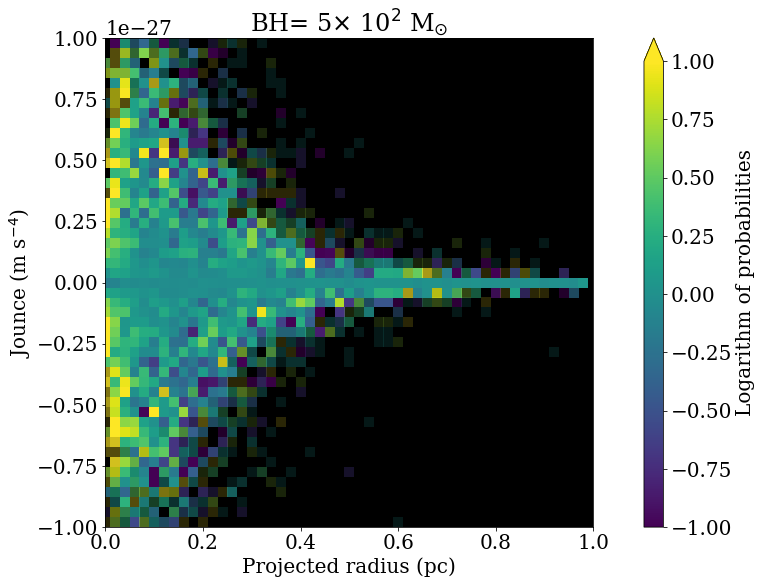}} %\quad
%\end{subfigure}

%\begin{subfigure}{.45\textwidth}
\subfloat{\includegraphics[height=0.29\textheight]{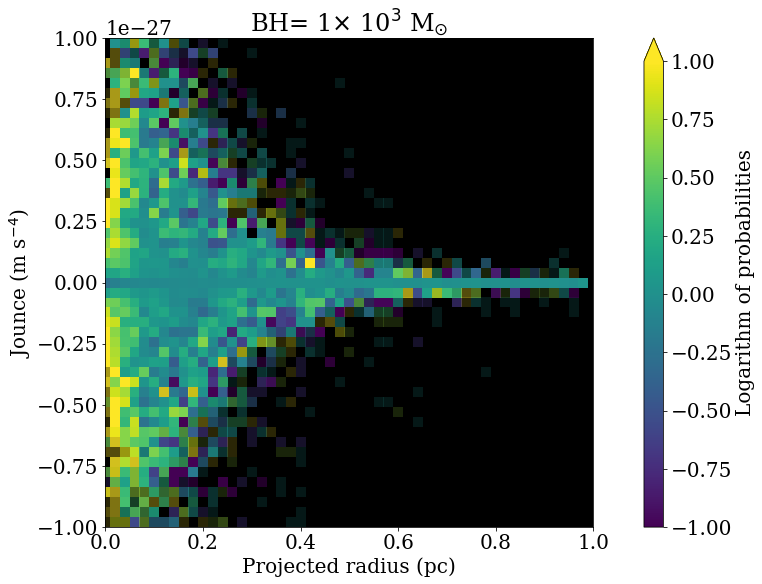}} %\quad
%\end{subfigure}
%\begin{subfigure}{.45\textwidth}
\subfloat{\includegraphics[height=0.29\textheight]{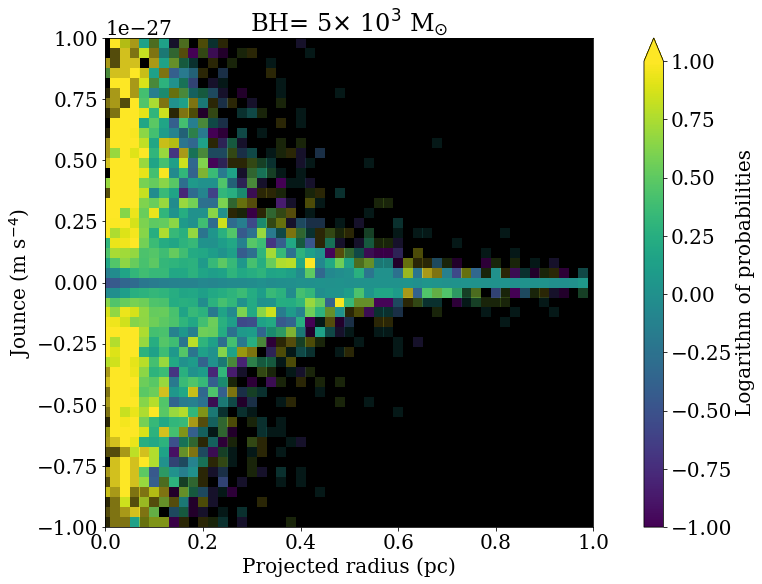}} %\quad
%\end{subfigure}

%\begin{subfigure}{.45\textwidth}
\subfloat{\includegraphics[height=0.29\textheight]{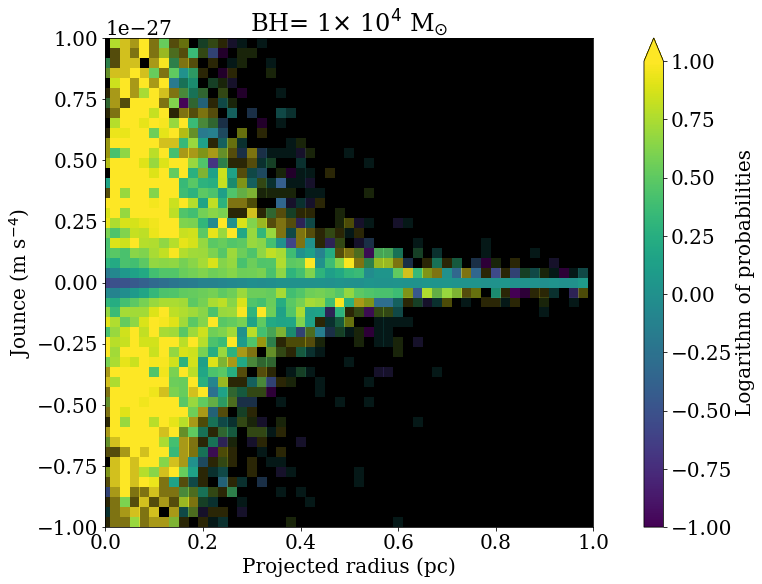}} %\quad
%\end{subfigure}
%\begin{subfigure}{.45\textwidth}
%\subfloat{\includegraphics[height=0.29\textheight]{new_jounce_prob_alpha_5e4.png}} %\quad
%%\end{subfigure}

\caption{Probability of finding an MSP in a given projected radius bin with a given jounce in the presence of an IMBH over the probability in the absence of it. The transparency of each pixel qualitatively shows the statistics of the value. Darker pixels have larger uncertainties. Pulsars found in a yellow square indicate that a black hole is likely present, while pulsars in blue square indicate that it is more likely that there is no black hole.}\label{jounce_prob_maps}
\end{figure*}

In this Section, we use MSPs as test particles for detecting the presence of an IMBH through their jerks and jounces. In order to determine the amount of information that can be derived from a single MSP, we map the different distributions in the jerk-radius plane for clusters with the IMBH and without. We repeat this process for all of the IMBH masses in our simulations. 
We compare the distributions by dividing the projected distance of the stars to the centre into 50 bins up to a maximum of 1 pc. The values of the jerks for these stars were divided into 50 bins from $-2\times 10^{-18}$ to $2\times 10^{-18}$ m s$^{-3}$. We show the comparison in Figure \ref{jerk_prob_maps}. The colour scale shows in a logarithmic scale the ratio of stars found in each cell in the case with an IMBH and in the case without. If an MSP is found in one of the bright yellow squares, then it is more likely that the parent cluster hosts an IMBH. On the contrary if an MSP is found in one of the dark blue squares it is more likely that it lives in a cluster without an IMBH. Figure \ref{jounce_prob_maps} shows the same probability maps for the jounces created dividing the jounces in 50 bins between $-10^{-27}$ and $10^{-27}$ m s$^{-4}$. These maps show 
that stars close to the centre of the cluster with high values of jerks and jounces are strong indicators of the presence of an IMBH while, if the jerks and jounces are small, the IMBH is likely absent. Even only a few stars located in the bright yellow or dark blue regions of the maps can help discriminate between the different IMBH masses.

We test whether the information gained from these maps is enough to infer the presence of an IMBH and its mass through Bayesian model selection. We assume that each cluster contains 20 MSPs for which accelerations, jerks and jounces have been measured with uncertainty equal to $10^{-21}$ m s$^{-3}$ for the jerks and to $10^{-29}$ m s$^{-4}$ for the jounces. Accelerations are measurable only for MSPs in binaries. The fraction of binary MSPs to isolated ones in the Galactic GCs varies form 0.1 for M15, to 0.5 for Terzan 5, and 0.6 for 47 Tuc and can reach 1 for M62. In the simulated cluster we assume that this fraction is 0.5. Therefore of the 20 MSPs assumed to be populating our clusters 10 are in a binary. Only for these 10 MSPs it is possible to measure the value of the line-of-sight acceleration. As a reference for the precision of this measurement we take the average uncertainty measured for the binary MSPs in 47 Tuc which is $10^{-9}$ m s$^{-2}$. We compare the results obtained using jerks and jounces with the ones obtained using only the accelerations.

Bayesian model selection is a statistical tool that allows us to compare quantitatively different models to find which one best describes the data. This comparison is based on the measures of the evidences which are the likelihoods of obtaining a specific value from the models integrated over the range of the free parameters. Since in our simulations there are no free parameters, the evidences correspond to the likelihoods, i. e. to the probabilities measured in each bin like in the maps of Figure \ref{jerk_prob_maps} and \ref{jounce_prob_maps} (the dimensions of the bins correspond to the errors of the acceleration, jerk and jounce). If there is no reason to prefer a priori one model over the other, the ratio of the evidences returns the Bayes factor which is the quantity used for the comparison. This quantity is usually reported in logarithmic values in base 10. If the Bayes factor is above 2 or below -2 (meaning that the evidence for one model is more than 100 times higher than the other) one model is strongly favoured. If it is between 1 and 2 or between -2 and -1 (with evidences between 10 and 100 times higher) one model is moderately favoured and if it is between -1 and 1 (evidences less than 10 times higher) nothing can be said.

For this test, we consider the simulations described earlier without the IMBH and extract 20 test particles representing MSPs. In Galactic GCs the radial distribution of pulsars does not reproduce the average radial distribution of stars. Since MSPs are more massive than the typical star in the cluster, they are more segregated in the centre and the radial distribution can be described by the following formula:

\begin{equation}\label{rad_distr}
N(R_{\perp}) = N_0 \left( 1 + \left(\frac{R_{\perp}}{r_{\rm c}}\right)^2 \right)^{\alpha/2},
\end{equation}
where $R_{\perp}$ is the projected distance from the cluster centre, $N_0$ is the normalisation and $\alpha$ is the spectral index responsible for the segregation. For the dominant mass stars in a cluster described by a King profile $\alpha= -2$. Typically for MSPs this index is $\sim -3$ \citep{Prager2017, Abbate2018}.  
To reproduce the observed distribution of MSPs, we extract their projected distance from the centre from equation  (\ref{rad_distr}).

%\red{Perche' tiri in ballo questa distirbuzione? Perche noi estraiamo le 20 MSP da questa distribuzione? Forse dovremmo dirlo esplicitamente... Inoltre, occhio perche non e' vero per i nostri clusters che le MSPs sono gli oggetti piu massicci... forse e' il caso di fare riferimento alla distribuzione, ma senza scrivere esplicitamente che "MSPs are more massive than the typical star in the cluster"}

We select the stars with projected distance within 0.02 pc of the position of the MSP and we extract the values of the line-of-sight acceleration, jerk and jounce to assign to the MSP from the values of the set of nearby stars. 
Then, we count the number of stars with projected distance within 0.02 pc of the MSP and with acceleration within the 1$\sigma$ interval of the value assigned to MSP, assumed to be $10^{-9}$ m s$^{-2}$. This number, divided by the total number of stars corresponds to the likelihood of a star to have that value of projected radius and line-of-sight acceleration. We multiply these likelihoods for each MSP together in order to obtain the total evidence for the set of MSPs. We repeat the same estimate for the jerks and the jounces using as uncertainties $10^{-21}$ m s$^{-3}$ and $10^{-29}$ m s$^{-4}$ respectively. 

We repeat the measure of the evidence for the same set of MSPs using the simulations with an IMBH described in Section \ref{simulations}. Finally, we divide the evidence in the case with an IMBH by the evidence in the case without to obtain the Bayes factor.
As the set of MSPs is small, this result is severely affected by the random extraction of the parameters. To gain a deeper insight in the result we repeat this calculation 2000 times for each simulation and average the results.

We compute the average Bayes factor for all IMBH masses considered in the simulations, first by taking accelerations into considerations, then jerks, and then jounces. Since accelerations, jerks and jounces are extracted independently of one another, the Bayes factors can be added together (when they are in logarithmic units). The results are summarised in Table \ref{tab_jerk_jounce}. Since the MSPs were extracted from the simulation without an IMBH the evidences will be higher in the case without an IMBH and the resulting Bayes factors are always negative.
At high masses (above $10^3$ $\msun{}$) the Bayes factor from the acceleration alone is below -2, so it is enough to affirm whether the IMBH is present. 
At a mass of $5\times 10^2$ $\msun{}$, the Bayes factor for the acceleration is only -1.6 so we need to incorporate the information from jerks and jounces in order to reach statistically reasonable levels of confidence. 
At the lowest black hole mass in the simulations ($10^2$ $\msun{}$) the total Bayes factor is -1 meaning that this mass is below our detection threshold. It is interesting however to look at how the single parameters contribute to this Bayes factor: the accelerations are barely informative about the presence of the IMBH with a Bayes factor of only -0.3 while the Bayes factor of the jerks is lower, -0.6. This shows that for black holes of low masses jerks can be more informative than accelerations. 
For all IMBH masses, jounces only contribute with a Bayes factor which is much less than that of accelerations or  jerks but still  jounces carry significant information which helps to bring the Bayes factor over the detection threshold.
From these results we state that in order to search for low mass IMBHs in GCs with MSPs, the best approach is to consider the accelerations and their derivatives together.

\begin{table*}
\caption{Logarithm of the Bayes factor between the model with an IMBH and without. The different black hole masses in units of M$_{\odot}$ and as percentage of the total mass of the simulated cluster are listed in column 1. Columns 2-4 give the Bayes factor using only acceleration, jerks and jounces and the last column give the total Bayes factor.
}

\begin{center}
\begin{tabular}[H]{c c c c c c}
\hline
{M$_{\rm BH}$ (M$_{\odot})$} & {Accelerations (log$_{10}$)} & {Jerks (log$_{10}$)} &  {Jounces (log$_{10}$)}  & {Total (log$_{10}$)}\\
\hline
\hline
$1 \times 10^4 \,(10\%)$ & -5.3 & -2.9 & -0.9 & -9.1\\
$5 \times 10^3 \,(5\%)$ & -4.4 & -2.3 & -0.7  & -7.4\\ 
$1 \times 10^3 \,(1\%)$ & -2.4 & -1.0 & -0.4 & -3.7\\
$5 \times 10^2 \,(0.5\%)$ & -1.6 & -0.8 & -0.3 & -2.7\\
$1 \times 10^2 \,(0.1\%)$ & -0.3 & -0.6 & -0.1 & -1.0\\
\hline

\end{tabular}\\
\label{tab_jerk_jounce}
\end{center}

\end{table*}

We test the limits of this technique by changing the number of MSPs in the clusters. We look for the number of MSPs necessary to detect an IMBH of $10^2$ $\msun{}$ in the simulations with statistical significance. This number (needed to retain a Bayes factor smaller than -2) is around 40 MSPs with 20 binaries. In this case the Bayes factor from the accelerations would be only -0.7 while the Bayes factor for the jerks would be -1.2 and for the jounces would be -0.2. 
If the number of MSPs is further reduced to 10 and of  binaries to 5, the limit mass of the IMBH measurable goes up to $10^3$ $\msun{}$. Also in this case the jerks and jounces play a significant role since the Bayes factor of the lone acceleration is -1.2, while combining the information from jerks and jounces we can reach -2.

\subsection{Application to 47 Tuc}\label{Section:47Tuc}

We apply the same technique to our $N$-body model of 47 Tuc.
The parameters of the MSPs were taken from the latest published ephemeris \citep{Ridolfi2016, Freire2017, Freire2018}. We used the observed projected distances from the centre, assuming a distance of the GC to the Sun of 4530 pc \citep{Bogdanov2016}. For binary MSPs we used the information on the period and orbital period derivatives to measure the acceleration and for all MSPs we used the $\ddot P$ to measure the jerk. For these pulsars there is still no clear measure of the jounces.
We repeat the analysis described in Section \ref{BH_pulsars}.  We first assumed that for isolated MSPs we don't have any information on the accelerations and then we repeat the analysis by estimating the accelerations by subtracting the inferred intrinsic spin-down from the period derivative. The intrinsic spin-down can be modelled by estimating the surface magnetic field inferred from Galactic MSPs with similar properties as those in 47 Tuc. 
%While it is true that the period derivative is heavily affected by the intrinsic spin down of the MSP, it is possible to model it after estimating the surface magnetic field inferred from Galactic MSPs with similar properties as those in 47 Tuc. 
A comprehensive list of Galactic MSPs with similar properties can be found in the Australian Telescope National Facility (ATNF) Pulsar Catalogue\footnote{ The full list of known MSPs and of their properties is available at: http://www.atnf.csiro.au/research/pulsar/psrcat/.} \citep{Manchester2005}. The probability distribution function is found to be log-normal with $\mu_{\log_{10}(B)}= 8.47$ and $\sigma_{\log_{10}(B)}= 0.33$ \citep{Prager2017}. For the pulsars in 47 Tuc the acceleration due to intrinsic spin-down is on average $\sim 5 \times 10^{-9}$ m s$^{-2}$. 

\begin{figure} 
\centering
\includegraphics[width=\columnwidth]{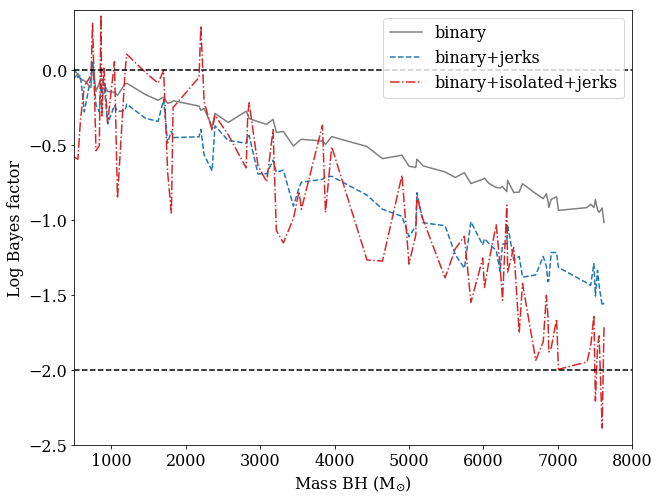}
\caption{Values of the Bayes factor for detecting a central IMBH in 47 Tuc using the parameters of acceleration and jerk measured from the MSPs dataset. The gray line considers only the acceleration from binaries. The dashed blue line considers the acceleration from the binaries and the jerks from all the MSPs. The dot-dashed red line considers also the acceleration of isolated MSPs for which the  value of the intrinsic spin-down is extracted from a probabilistic distribution.
}
\label{Bayes_factor_now}
\end{figure}

The results of the analysis with and without the acceleration from isolated MSPs are shown in Figure \ref{Bayes_factor_now} where we give the Bayes factor for different IMBH masses. We first show in grey the Bayes factor obtained using only the acceleration from binary MSPs. In this case, high mass black holes look disfavoured but there is not enough statistics to exclude any of the tested IMBH mass. The dashed blue line represents the Bayes factor measured adding the information from the jerks. The situation improves but the Bayes factor never reaches the value of -2. Only when combining also the acceleration from the isolated MSPs,  we find a statistically significant upper limit to the mass of the IMBH. This limit is $\sim 7000$ $\msun{}$. The accelerations for isolated MSPs are measured as described above and add a lot of fluctuations due to the unknown intrinsic spin-down.

The upper limit we obtain here is not as stringent as in \cite{Abbate2018} where only the accelerations of the MSPs were used. This is because \cite{Abbate2018} adopted a Monte Carlo Markov Chain algorithm to determine the 3D position of the MSPs. The determination of these positions allowed the authors to estimate analytically the expected line-of-sight accelerations in presence of black holes of different masses. 
%In this work, this approach is not feasible since we lack of analytical or statistical expressions to calculate the contribution to the jerks due to nearby neighbours in presence of an IMBH. However, the focus of the paper is to show the improvement brought by jerks 
%over the accelerations in the search for a central IMBH. In order to do a fair comparison between the way we treated accelerations and jerks, we decided not consider the 3D positions of the MSPs for either quantities.
 This approach is not feasible for the jerks since we lack the analytical or statistical expressions to calculate the contribution due to nearby neighbours in presence of an IMBH. Therefore, it is not possible to estimate the total jerk felt by a star given the position along the line of sight. However, the focus of the paper is to show the improvement brought by jerks over the accelerations in the search for a central IMBH. In order to do a fair comparison between the accelerations and the jerks, we decided to disregard the 3D positions of the MSPs when measuring the accelerations.

\subsection{Improvement with future observations}

Future observations can lead to a major improvement to the mass measurement of the IMBH.
The uncertainty with which the acceleration can be measured decreases with longer observing baselines as $T^{-5/2}$ with $T$ being the total observing timescale \cite{Liu2018}. In contrast, the uncertainty on the jerks decreases as $T^{-7/2}$. The uncertainty on the jounces decreases even faster as $T^{-9/2}$ leading to a possible measure of jounces for all MSPs in 47 Tuc in the near future. As the uncertainties on the jerks and jounces of MSPs decrease faster than the ones on the accelerations, a technique to extract information from the jerks and jounces will be necessary to be more sensitive to lower IMBH masses.

To test how the improvements of the measurements of jerks and a first measurement of jounces would influence the sensitivity of the search of a central IMBH, we performed mock simulations of future observations. We used the software {\scriptsize TEMPO2} \citep{Edwards2006} to simulate the ToAs of the pulses supposing that the cluster will be observed on a monthly basis for five more years with the same sensitivity as up to now. 
With the resulting reduced uncertainties for the jerks and jounces, we apply the same technique explained in Section \ref{BH_pulsars} again for 47 Tuc. 
%First, we extract values of jerks and jounces from the simulations without a black hole at the radius corresponding to each pulsar and then we compare the number of stars with jerk and jounce within the uncertainty range in the case of simulations without and with a black hole. The mass of the black hole that reaches a logarithm of the Bayes factor less then -2 is the mass limit that we could be able to probe with the refined data.

\begin{figure} 
\centering
\includegraphics[width=\columnwidth]{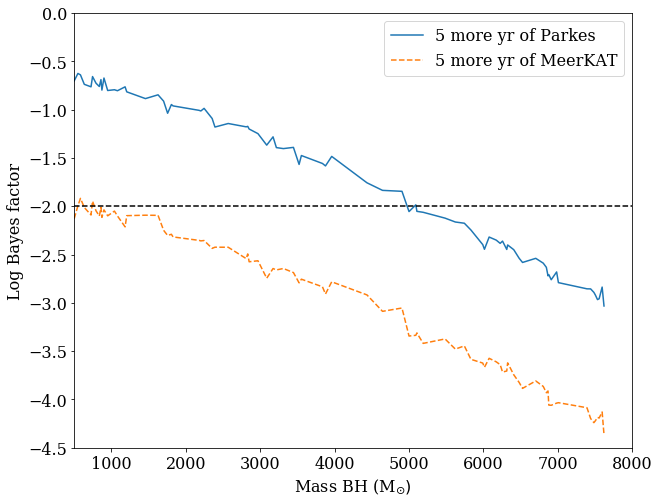}
\caption{Prediction of detectability of a central IMBH in 47 Tuc assuming we have access to 5 more years of timing observations with the Parkes radio telescope (shown in blue) and with the MeerKAT radio telescope (shown in dashed orange). The simulations show that Parkes would allow us to detect IMBHs of mass $> 5000$ M$_{\odot}$ and that with MeerKAT of mass $> 1000$ M$_{\odot}$. Details of how these simulations were run are found in the text.}
\label{Bayes_factor_pks_mk}
\end{figure}

The blue line in Figure \ref{Bayes_factor_pks_mk} shows the Bayes factor as a function of IMBH mass if observations are performed with the Parkes radio telescope. The IMBH upper limit in this case reduces to $\sim 5000$ M$_{\odot}$.
%The pulsars in the cluster have been observed with the Parkes radio telescope in Australia. As is shown in Figure \ref{Bayes_factor_pks_mk}, if we observe the pulsars on a monthly basis over five years at Parkes we could in principle detect a black hole of $\sim 5000$ M$_{\odot}$. 

We repeat the same simulations for the MeerKAT radio telescope in South Africa. We first estimate the increase in sensitivity this telescope would have compared with Parkes. MeerKAT is made up of 64 antennas of 13.5 m of diameter, the nominal gain of this telescope is $G_{\rm M} \sim 2.8$ K Jy$^{-1}$ with an observing bandwidth of $\Delta\nu_{\rm M}=856$ MHz. The gain of Parkes instead is $G_{\rm P}=0.64$ K Jy$^{-1}$ and the bandwidth used for these observations is $\Delta\nu_{\rm P}=256$ MHz. The signal to noise ratio of an observation with MeerKAT would be $G_{\rm M}/G_{\rm P} \sqrt{\Delta\nu_{\rm M}/\Delta\nu_{\rm P}} \sim 8$ times higher. This roughly translates into a timing precision with MeerKAT observations 8 times higher than with Parkes. With these improved sensitivities we expect to detect in five years IMBHs of $\rm M_{\rm BH} > 1000$ M$_{\odot}$ as shown in the dashed orange line in Figure \ref{Bayes_factor_pks_mk}. 

Figure \ref{Bayes_factor_pks_mk} shows not only the upper limits for non-detections of IMBHs but also the lowest detectable mass of IMBHs in future observations. Any IMBH of mass higher than the upper limits determined above will be detected as the accelerations, jerks and jounces of the MSPs will be inconsistent with the simulation without the IMBH.

\subsection{Caveat}

%When discussing these simulations, an important point to emphasize is how the density of stars is influenced by the IMBH. 
In Section \ref{section_jerks} we introduced the concept of an over-density of stars inside the influence radius of the IMBH described by a power law \citep{Baumgardt2004b}. This over-density is a stable configuration attained over a large number of relaxation times. Our simulations are not run for such long times to recover the stationary profile, but for the mock cluster used, the relaxation time \citep{Binney2008} measured at the core radius  is close to the duration of the simulations, i. e. 50 Myr. For this reason we see an enhanced density in the centre that helps increasing jerks and jounces inside the influence radius. 

The situation differs for 47 Tuc. In the simulations we don't see the over-density forming in the centre as the core relaxation time is much longer, $\sim 700$ Myr. Since the density does not change, the enhanced jerks and jounces that we measured are only caused directly by the gravitational potential of the IMBH and increased velocity dispersion.
As jerks and jounces increase in presence of the over-density (as shown in the mock cluster simulations) the results derived in Section \ref{Section:47Tuc} highlighting the importance of jerks in the search of an IMBH will be strengthened by using more evolved simulations.
How jerks and jounces are affected by an evolving density profile
until relaxation is completed  will be addressed in a forthcoming paper.
%and will b. Detailed simulations until complete relaxation is reached and sampled at different evolutionary stages of the cluster are needed to answer this question.

\section{Conclusions} \label{conclusion}

IMBHs in Galactic GCs have been searched for extensively, and compelling evidence is still lacking. In particular the discovery of IMBHs of mass below a few thousands solar masses is challenging since their influence is limited to the centre-most region of the star cluster. MSPs have been considered as key probes through their accelerations measured via pulsar timing analysis.
In this paper we show that measurements of the second and third derivative of the rotational period of an ensemble of MSPs in a Galactic GCs help in their identification. These derivatives are correlated respectively with the line-of-sight component of the first and second derivatives of the acceleration called jerks and jounces. 

Direct $N$-Body simulations of star clusters using the Hermite's 6th order integrator require the computation of higher order derivatives of the star's accelerations to trace with high accuracy the stellar dynamics determined by the mean gravitational  field and by neighbouring stars.  For the first time, we read off the values of the star's jerks and jounces from a suite of simulations with and without IMBHs.

We demonstrated that a central IMBH modifies the distribution of jerks and jounces within its sphere of influence.
We used MSPs as test particles of the gravitational field and, using a Bayesian analysis, we 
computed the probability of finding MSPs in a given jerk and jounce bin extracted in a suite of simulations with IMBHs over a range of mass between $10^2$ and $10^4$ $\msun{}$.
%We showed that these effects are measurable with high accuracy using MSP timing techniques given a long enough observational baseline. 
%We applied a very simple technique based on comparison of these measures with simulations of the clusters with different black holes. 
We showed that the derivatives of the acceleration are more sensitive to the presence of black holes of low masses when compared with just the information taken from the acceleration. A combination of all the kinematic information is shown to be the best method for searching for black holes of all sizes.

We applied this technique to the MSPs in 47 Tuc and obtained an upper limit on the mass of the central IMBH of $\sim 7000$ M$_{\odot}$. 
This result is not as stringent as other upper limits derived for this cluster based only on the acceleration data of the MSPs \citep{Abbate2018} because it does not take into consideration the position along the line of sight of the MSPs. To include this parameter we need a description of the effects of the IMBH on the jerks and jounces caused by neighbouring stars which is still lacking. However, the inclusion of jerks in the calculation was essential in measuring this mass limit.

%This technique cannot test masses as low as previous results based only the acceleration of the pulsars \citep{Kiziltan2017a,Abbate2018}. This is due to the simplicity of the technique used and to the fact that we did not estimate the position along the line of sight of the pulsars. However, the inclusion of the jerks in the calculation was essential in measuring this mass limit. A more detailed modeling that includes an probabilistic description of the effects of the black hole on the jerks and jounces caused by nearby stars would be needed to probe black holes of lower masses.

As more observations pile up, the accuracy of the measurements of jerks and jounces improves faster than for accelerations. The uncertainty scales as $T^{-7/2}$ for the jerks and $T^{-9/2}$ for the jounces where $T$ being the total length of observations. In contrast the accuracy of accelerations in the case of binary pulsars scales as $T^{-5/2}$. For isolated pulsars the accuracy of the acceleration will not increase since the intrinsic spin down is unknown. 

In the near future jerks and jounces might be so precise that they are more informative than accelerations. 
This has been tested simulating observations for five years with the Parkes radio telescope and with the MeerKAT radio telescope which will observe this cluster in the future. We estimated how the measurements of jerks and jounces can improve and applied the technique to set lower upper limits for IMBHs. With five more years of observations at Parkes the upper limit is $\sim 5000$ M$_{\odot}$. With MeerKAT the limit is below 1000 M$_{\odot}$.

MSPs can become the most precise probes to search for the presence of IMBHs in Galactic GCs. Long observational campaigns are necessary to reach the precision required to achieve this result. Thus, it is important that all these MSPs, even the isolated ones, are regularly observed. Newly discovered MSPs close to the centre of the clusters would also greatly increase the accuracy of the search.

\section*{Acknowledgments}
We thank Andrea Possenti for providing insight on simulations of future pulsar observations and Holger Baumgardt for a critical reading of the manuscript and useful comments. We thank the anonymous referee for the helpful comments.
We acknowledge the ``Accordo Quadro INAF-CINECA (2017)'' and the CINECA-INFN agreement for the availability of high performance computing resources and support. MS acknowledges funding from the European Union's Horizon 2020 research and innovation programme under the Marie-Sklodowska-Curie grant agreement No. 794393.
MC  acknowledges funding from the INFN experiment Teongrav.

\bibliographystyle{mnras}
\bibliography{Jerks_bib}

\appendix

\section{Tests of the equations for jerks and jounces}\label{appendix}

\begin{figure*}
\centering
%\begin{subfigure}{.45\textwidth}
\subfloat {\includegraphics[width=\columnwidth, height=0.3\textheight]{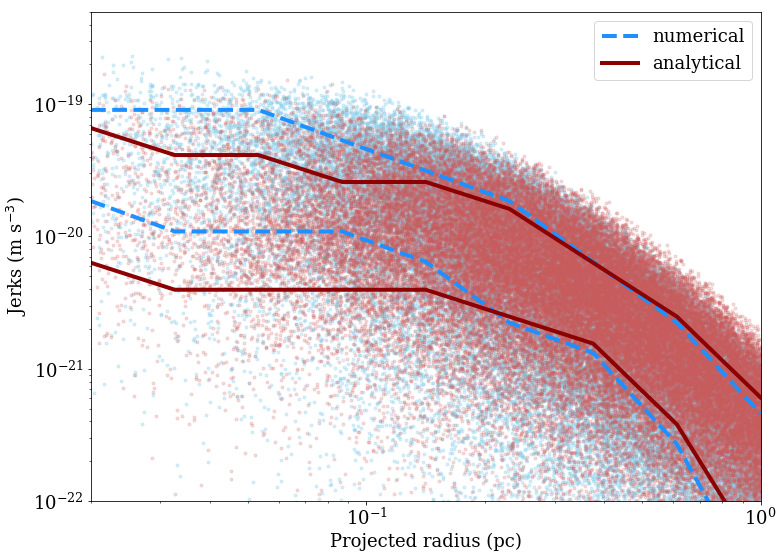}} %\begin{subfigure}{.45\textwidth}
\subfloat{\includegraphics[width=\columnwidth, height=0.3\textheight]{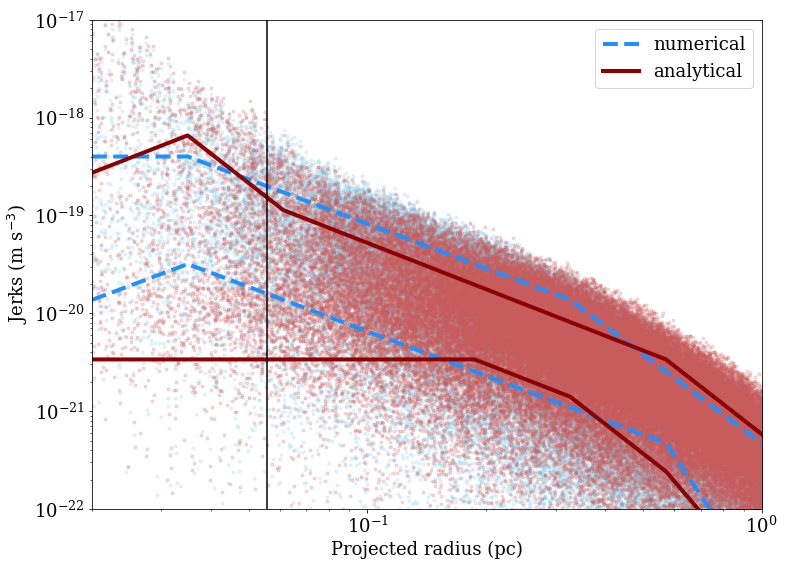} }%\quad
\caption{Values of the mean field jerks measured in the case without (left panel) and with an IMBH of 1000 $\msun{}$ (right panel). Blue dots represent the mean field jerks measured numerically from the simulations while red dots are the results of the analytical formulae. The dashed blue and the solid red lines bracket the 1$\sigma$ interval of the distributions. In the right panel the vertical line shows the influence radius of the IMBH.
}\label{jerk_analytic}
\end{figure*}

\begin{figure*}
\centering
%\begin{subfigure}{.45\textwidth}
\subfloat {\includegraphics[width=\columnwidth, height=0.3\textheight]{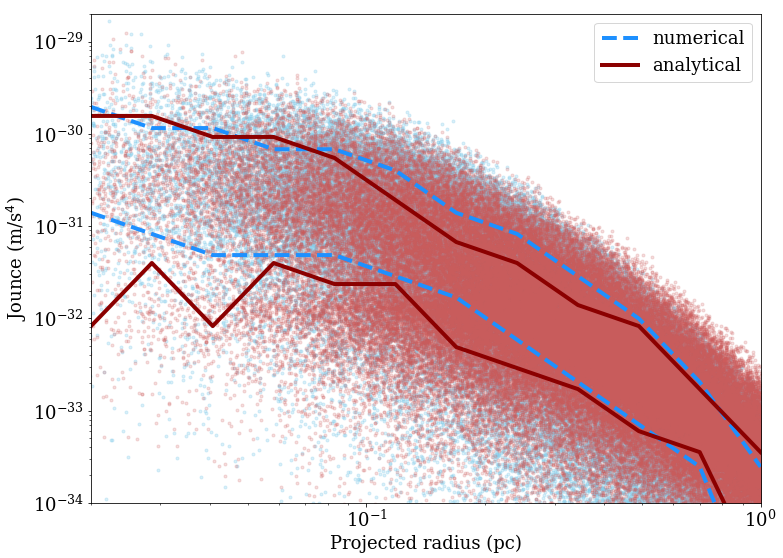}} %\begin{subfigure}{.45\textwidth}
\subfloat{\includegraphics[width=\columnwidth, height=0.3\textheight]{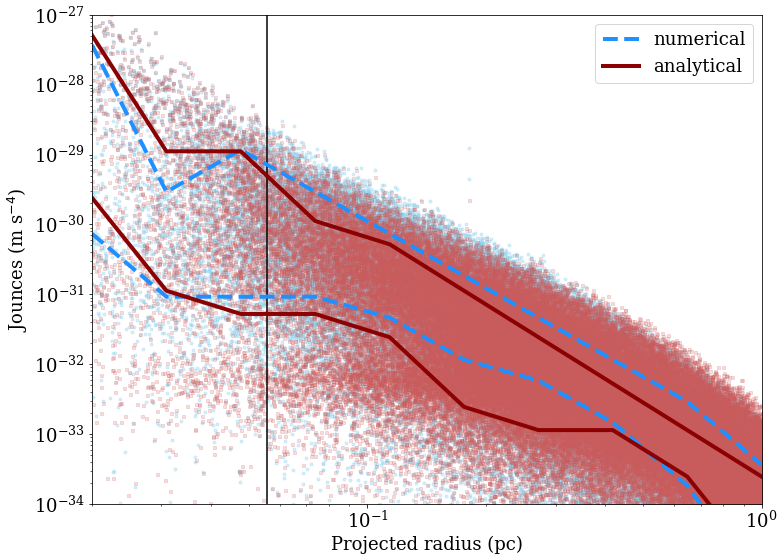} }%\quad
\caption{Values of the mean field jounces measured in the case without (left panel) and with an IMBH  of 1000 $\msun{}$(right panel). The blue dots represent the mean field jounces measured numerically from the simulations while the red dots are the results of the analytical formulae. The dashed blue and the solid red lines bracket the 1$\sigma$ interval of the distributions. In the right panel the vertical line shows the influence radius of the IMBH.
}\label{jounce_analytic}
\end{figure*}
In this appendix we compare the formulae for the mean field jerks and jounces in GCs derived in Section \ref{section_jerks} and \ref{section_jounces} and those extracted in simulations of the mock cluster both in the case with and without a central IMBH. 

The left panel of Figure \ref{jerk_analytic} 
shows the case of jerks in the simulation without an IMBH. Red dots refer to the stellar jerks calculated analytically using equation (\ref{jerk_king}) while the blue dots show the mean field stellar jerks extracted numerically from the simulation by estimating the mass enclosed within the distance from the centre of each star and using equation (\ref{jerk_general}) and (\ref{jounce_general}).
The jerks calculated analytically are of the same order of magnitude of the ones measured numerically but they appear to be smaller in the centre of the cluster. This discrepancy is caused by stars that are close the centre of the cluster in projection on the plane of the sky but are distant along the line of sight. The analytical equations for stars more distant than a few core radii systematically underestimate the value of the jerks when compared with the numerical value. The right panel of Figure \ref{jerk_analytic} shows the case of a simulation with a central IMBH of 1000 M$_{\odot}$. We use equation  (\ref{jerk_general}) and (\ref{jerks_cusp}) to estimate analytically the jerks. In this case, the analytic estimate appears to be consistent with the numerical one. Inside the sphere of influence of the IMBH the analytical distribution extends to lower values of the jerks because of the same projection effect as in the case without an IMBH.

The left panel of Figure \ref{jounce_analytic} shows the jounces calculated in a simulation without an IMBH. Red dots show the mean field stellar jounces measured using equation (\ref{jounce_king}) while blue dots show jounces computed numerically. The two distributions are compatible with one another showing that equation (\ref{jounce_king}) is good analytical approximation of the mean field jounces. The same can be said for the equations describing the effects of the IMBH as is seen on the right panel again for a 1000 M$_{\odot}$ IMBH. In the analytical computation, the IMBH contribution is inferred from equation (\ref{jounce_general}) and for the stellar cusp using equation (\ref{jounce_cusp_1}) and (\ref{jounce_cusp_2}). 
In both cases with and without the IMBH the clustering of stars with low jounces at small projected distances is caused by stars that are distant from the centre but appear close in the projection along the line of sight. For these stars the analytical formula is not valid and tends to underestimate the jounces.

% Don't change these lines
\bsp	% typesetting comment
\label{lastpage}
\end{document}